\documentclass[12pt]{article}
\usepackage[centertags]{amsmath}
\usepackage{amsfonts}
\usepackage{amssymb}
\usepackage{amsthm}
\usepackage{newlfont}
\usepackage{epsfig}
\usepackage{amscd}

\newcommand{\RR}{{\mathbb R}}
\newcommand{\ZZ}{{\mathbb Z}}

\newcommand{\CC}{{\mathbb C}}
\newcommand{\beq}{\begin{equation}}
\newcommand{\eeq}{\end{equation}}
\newcommand{\ba}{\begin{array}}
\newcommand{\ea}{\end{array}}
\newcommand{\bea}{\begin{eqnarray}}
\newcommand{\eea}{\end{eqnarray}}
\renewcommand{\Re}{\mathop{\mathrm{Re}}}
\renewcommand{\Im}{\mathop{\mathrm{Im}}}

\input{tcilatex}

\begin{document}

\begin{center}

{\Large\bf Newtonian dynamics in the plane \\
corresponding to straight and cyclic motions on the hyperelliptic curve
$\mu^2=\nu^n-1,~n\in\ZZ$:\\
ergodicity, isochrony, periodicity and fractals }


\bigskip

\textbf{P. G. Grinevich}$^{1,a}$\textbf{and P. M. Santini}$^{2,b}$.

\medskip

$^{1}$ Landau Institute for Theoretical Physics, Moscow, Russia

$^{2}$ Dipartimento di Fisica, Universit\`{a} di Roma "La Sapienza" and \\

Istituto Nazionale di Fisica Nucleare, Sezione di Roma

\bigskip

$^{a}$pgg@itp.ac.ru

$^{b}$paolo.santini@roma1.infn.it

\bigskip

\end{center}

\begin{abstract}
We study the complexification of the one-dimensional Newtonian
particle in a monomial potential. We discuss two classes of motions on the
associated Riemann surface: the rectilinear and the cyclic motions,
corresponding to two different classes of real and autonomous Newtonian dynamics
in the plane. The rectilinear
motion has been studied in a number of papers, while the cyclic motion is much less
understood. For small data, the cyclic time trajectories lead to isochronous dynamics.
For bigger data the situation is quite complicated; computer experiments show that,
for sufficiently small degree of the monomial,
the motion is generically periodic with integer period, which depends in a quite
sensitive way on the initial data. If the degree of the monomial is sufficiently
high, computer experiments show essentially chaotic behaviour. We suggest a possible
theoretical explanation of these different behaviours. We also introduce a one-parameter
family of 2-dimensional mappings, describing the motion of the center of the circle,
as a convenient representation of the cyclic dynamics; we call such mapping the center map.
Computer experiments for the center map show a typical multi-fractal behaviour with
periodicity islands. Therefore the above complexification procedure generates
dynamics amenable to analytic treatment and possessing a high degree of complexity.
\end{abstract}
\newpage

\section{From Newtonian systems to the motion on Riemann surfaces}

In this paper we investigate the following two classes of autonomous Newtonian
dynamical systems in the plane
\beq
\label{ODEs12}
\ba{l}
\ddot{\bf x}={\bf F^{(m)}_1}({\bf x}),~~~m\in\ZZ ,~~m\ne -1,\\
\ddot{\bf x}={\bf F^{(m)}_2}({\bf x},\dot{\bf x}),~~~m\in\ZZ,~~m\ne -1,
\ea
\eeq
where ${\bf x}=(x,y)\in\RR^2$, and $\ddot f=d^2f/dt^2$. The forces $\bf F^{(m)}_{1,2}$ are defined by
\beq
\label{Forces}
\ba{l}
{\bf F^{(m)}_1}({\bf x})\equiv -\left(Re(\alpha^2(x+iy)^m),Im(\alpha^2(x+iy)^m))\right), \\
{\bf F^{(m)}_2}({\bf x},\dot{\bf x})\equiv -\dot{\bf x}\wedge{\bf h}_m-8\pi^2\frac{m+1}{(m-1)^2}{\bf x}-
\left(Re(x+iy)^m,Im(x+iy)^m\right),
\ea
\eeq
where the constant vectors ${\bf h}_m,~m\in\ZZ$ are orthogonal to the $(x,y)$ plane, with
$||{\bf h}_m||=2\pi\frac{m+3}{m-1}$, and $\alpha\in\CC$ is an arbitrary
parameter.

The classes (\ref{ODEs12}a) and (\ref{ODEs12}b) originate from the same complex ODE:
\begin{equation}
\label{unmodified}
\frac{d^2\zeta}{d\tau^2}=-\zeta ^m~,~~m\in\ZZ,~~m\ne -1,~~\tau\in\CC,~~\zeta \equiv \zeta \left( \tau
\right)\in\CC ~,
\end{equation}
solved by the complex quadrature
\beq
\label{quadrature1}
\ba{l}
\tau=\int\limits_{\zeta(0)}^{\zeta}\frac{d\zeta}{\sqrt{2(E-\frac{\zeta^{n}}{n})}}, \\
E=\frac{1}{2}\left(\frac{d\zeta}{d\tau}(0)\right)^2+\frac{1}{n}(\zeta(0))^n, \ \ \ n= m+1 .
\ea
\eeq
The first class (\ref{ODEs12}a) is directly obtained from (\ref{unmodified}) via the following change
of independent variables
\beq
\label{straight}
\ba{l}
z(t)=x(t)+iy(t)=\zeta(\tau), \\
\tau(t)=\alpha t,~~~\alpha\in\CC
\ea
\eeq
corresponding to a rectilinear motion on the hyperelliptic curve of the quadrature (\ref{quadrature1});
the second class (\ref{ODEs12}b) is instead connected to (\ref{unmodified}) via the following change of
dependent and independent variables (Calogero's transformation) \cite{Calogero1}:
\beq
\label{trick}
\ba{l}
z\left( t\right) =x(t)+iy(t)=\exp \left( \frac{i\,4\,\pi\,t}{n-2}\right) \,\zeta
\left( \tau \right) ~,  \\
\tau(t) =\frac{\exp (i\,2\,\pi\,t)-1}{2\pi i}.
\ea
\eeq
Equation (\ref{trick}b) implies that, as the ``physical time'' $t$ travels onward along the real
$t$-axis, the motion of the complex time $\tau$ is cyclic; the auxiliary transformation (\ref{trick}a)
is what makes the Newtonian class (\ref{ODEs12}b) autonomous.

Through the change of variables
\beq
\label{change}
\ba{l}
w(\xi)=c_1\zeta(\tau),\;\;\;\xi=c_2\tau +\xi_0, \\
c_1=\left(\frac{1}{nE}\right)^{\frac{1}{n}},~~~~c_2=\sqrt{2}\left(E\right)^{\frac{n-2}{2n}}
\left(\frac{1}{n}\right)^{\frac{1}{n}}, \ \ \
\xi_0=\int\limits_{0}^{c_1\zeta(0)}\frac{dw}{\sqrt{1-{w}^n}},
\ea
\eeq
the quadrature (\ref{quadrature1}) takes its adimensional form.
We have 2 cases, corresponding to $n$ positive and $n$ negative respectively:
\beq
\label{quadrature2}
\xi=\int\limits_0^{w}\omega,
\eeq
\beq
\label{quadrature3}
\omega=\frac{dw}{\sqrt{1-{w}^n}}, \ \ \ n>0; \ \ \ \ \ \
\omega= \sqrt{\frac{w^{|n|}} {{w}^{|n|}-1}} dw, \ \ \ n<0.
\eeq

Therefore the Newtonian systems (\ref{ODEs12}a) and (\ref{ODEs12}b) correspond, via (\ref{straight}),
(\ref{trick}), (\ref{change}), respectively  to the rectilinear motion in the $\xi$-plane:
\beq
\label{xi-straight}
\xi = \xi_0 + \alpha c_2 t
\eeq
and to the cyclic motion in the $\xi$-plane:
\beq
\label{xi-cyclic}
\ba{l}
\xi = \xi_c + R\exp{(i(2\pi t+\varphi))}, \\
\xi_c=\xi_0-\frac{c_2}{2\pi i}, \ \ \ R=\frac{|c_2|}{2\pi}, \ \ \ \varphi=\mbox{arg}(-ic_2),
\ea
\eeq
where $\xi_c$ and $R$ are the center and the radius of the corresponding circle. We remark that
the Cauchy data of the Newtonian dynamics (\ref{ODEs12}b) determine completely
this circle and the initial point on it.

To express $w$ as a function of $t$ we have to invert the hyperelliptic integrals
(\ref{quadrature2}), (\ref{quadrature3}) defined on the surfaces
\beq
\Gamma: \mu^2 = 1-w^{|n|}, \ \  n>0 \ \mbox{or}\  n=2k<0,
\eeq
\beq
\Gamma_1: \mu^2 = w-w^{|n|+1}, \ \  n=2k+1<0;
\eeq
therefore the trivial dynamics (\ref{xi-straight}) and (\ref{xi-cyclic}) in the
$\xi$-plane result in  highly nontrivial motions on $\Gamma$, $\Gamma_1$.

Let us point out that, for $n>0$, we have the problem of inverting a hyperelliptic
integral defined by a {\bf holomorphic} 1-form, and for $n<0$ we have to invert
an integral defined by a {\bf meromorphic} 1-form. As we shall see below, if the
 motion is rectilinear, the case $n<0$ is much simpler from the dynamical point of view.

The rectilinear motion on Riemann surfaces can be naturally
interpreted as Hamiltonian system with a multivalued Hamiltonian (
or, equivalently, as a Hamiltonian system defined by a Hamiltonian
1-form with non-zero periods, see review \cite{Nov1}). This motion
was intensively studied in the literature (references can be found
in the reviews \cite{Z1,DN}).  A new description of the topology of these flows was suggested
recently in \cite{Nov2}.

The interest to this problem was partially motivated by its
connections to stochastic dynamics. Consider a cycle on the Riemann surface
transversal to the flow. The Poincar\'e map on this
cycle preserves the measure on the cycle, therefore it can be interpreted as
the interval exchange map, which is one of the basic stochastic models.
Moreover, any interval exchange map can be obtained from this construction.
It turns out that the properties of this map are deeply connected with the
geometry of the moduli spaces \cite{Konts, Masur, Veech, Z1, Z2, Z3}.
In particular, the Lyapunov exponents for this
map at generic points can be calculated in terms of the Lyapunov exponents for
the geodesic flow on the Teichmuller space.

Another motivation for the study of rectilinear motions
comes from solid state physics: the conductivity
tensor for metals in a ``strong enough'' magnetic fields is deeply connected to
the quasiclassicl electron motion on the Fermi-surface, which is a Hamiltonian flow
on the Fermi-surface, and the Hamiltonian 1-form is determined by the direction
of the magnetic field \cite{MN}, \cite{DN}. In contrast to the case of the interval exchange map,
the holomorphic form is now non-generic, and its real part has exactly 3 non-zero
periods; therefore this situation needs a separate study.

The study of cyclic motions on Riemann surfaces was initiated in \cite{Calogero1}, where a
general procedure, involving the change of variable (\ref{trick}), was introduced
to deform large classes of ODEs so that they feature a lot of periodic solutions.
The deformed dynamical systems are isochronous for initial
data lying in a certain domain of full dimensionality (corresponding to the situation in which
the circle is contained inside the analyticity disc of the unmodified ODE).
Outside this domain, the circle may include branch points; then
bifurcations occur, and either the motion becomes periodic with multiple period,
or the periodicity is lost and the system may exhibit sensitive dependence on the
initial data (see \cite{Calogero1, Calogero2, CF, CS,CFS, Calogero3, Calogero4, CGSS}
and references therein quoted). This
mathematical mechanism was first described in \cite{CS,CFS}.
As it was shown in \cite{CGSS} for an aristotelian 3-body problem which allows for a
quite explicit analytical treatment, the transition from regular (periodic) to irregular
(with sensitive dependence on the initial data) dynamics takes place when an infinity
of branch points of the solution are arbitrarily close to the real $t$-axis.

A very general class of autonomous dynamical systems obtained through Calogero's transformation
has been constructed in \cite{Calogero5} (see also \cite{Induti}). This class contains, in particular,
systems of ODEs
solved by quadratures associated with elliptic and hyperelliptic curves; the first
investigation of some examples of this type was presented in
\cite{Induti}. Also the Newtonian systems (\ref{ODEs12}b), corresponding to
cyclic motions on the Riemann surface of the quadratures (\ref{quadrature1}), are examples
belonging to this general class. A detailed study, analytical and computational, of the
systems (\ref{ODEs12}b) in the particular cases $n=4,5,6$  has been recently done in \cite{FG},
with the following results.
In the elliptic case $n=4$ the motions are isochronous with period $1$; if $n=6$ the
hyperelliptic curve doubly covers an elliptic curve, and all motions are
periodic with periods $1$ or $2$. In the more interesting case $n=5$, numerical experiments
seem to indicate that all solutions are periodic, with arbitrarily large
period. The results contained in \cite{Calogero5,Induti,FG} have been the main motivation
of our study.

\vskip 5pt
\noindent
{\bf Remark 1.} The cyclic motion can be described as the motion with constant
geodesic curvature on the Riemann surface with almost everywhere flat singular metric (or,
equivalently, as the motion with a constant geodesic curvature on a polyhedron).
As it was pointed out by S.P.Novikov \cite{Nov1}, in the 2-dimensional case the motion with
a prescribed geodesic curvature can be naturally interpreted as the Hamiltonian motion in
magnetic field.

\vskip 5pt
\noindent
{\bf Remark 2.} Under the change of variables (\ref{straight}), the complex Newtonian equation
(\ref{unmodified})
becomes $\ddot z=-\alpha^2 z^m$, and can be cast into the complex Hamiltonian form:
\beq
\ba{l}\dot z=\frac{\partial {\cal H}}{\partial \pi}, \ \ \ \
\dot \pi=-\frac{\partial {\cal H}}{\partial z}, \ \ \ z,\pi\in \CC, \\
{\cal H}(z,\pi)\equiv \frac{\pi^2}{2}+\frac{\alpha^2}{n}z^n, \ \ \pi=\dot z.
\ea
\eeq
Then it follows that also the corresponding dynamical systems (\ref{ODEs12}a)
can be written in Hamiltonian form:
\beq
\label{realHam}
\ba{l}
\dot x=\frac{\partial H}{\partial p_x},\ \ \ \ \ \ \dot y=\frac{\partial H}{\partial p_y}, \\
\dot p_x=-\frac{\partial H}{\partial x},\ \ \ \ \ \ \dot p_y=-\frac{\partial H}{\partial y},
\ea
\eeq
where the canonically conjugated real variables $\vec x=(x,y)$ and $\vec p=(p_x,p_y)$  are defined by
$z=x+iy, ~\pi=p_x-ip_y$, and the
Hamiltonian $H(\vec x,\vec p)$ is the real part of $\cal H$ (the general procedure to go from complex to
real Hamiltonian formulations can be found in \cite{Strocchi}). In addition,
the real and imaginary parts of ${\cal H}$ are constants of motion in
involution for the Hamiltonian system (\ref{realHam}). Therefore the Newtonian equations
(\ref{ODEs12}a) are Liouville integrable, with polynomial (if $n>0$)
or rational (if $n<0$) constants of motion in involution. But this Liouville integrability
{\bf is not global}. Indeed, a simple local analysis shows
that the singularities of $\vec x(t)$ and $\vec p(t)$ in the complex $t$ - plane are, for $n\ne 3,4$,
branchpoints (see (\ref{BPs})). In addition, for $n\ne 6$,
such branch points are everywhere dense in the complex $t$ - plane, $\vec x,~\vec p$ are unbounded
and the 2-dimensional real variety defined by the two constants of motion is not a torus
(in particular, global action - angle variables cannot be defined).
The existence of everywhere dense branch point singularities, which causes the
lack of global integrability in the Liouville sense, turns out to be the reason for the sensitive
dependence on the initial conditions exhibited by the corresponding dynamics, discussed in the
following sections.

For the Newtonian class (\ref{ODEs12}b), which clearly inherits from (\ref{unmodified}) and (\ref{trick}) the
two $t$-dependent first integrals $\tilde H$, $\tilde K$:
\beq
\ba{l}
\tilde H=\mbox{Re}~\tilde{\cal H}, \ \ \ \ \tilde K=\mbox{Im}~\tilde{\cal H}, \\
\tilde{\cal H}\equiv \left[ \frac{1}{2}(\dot z-i\frac{4\pi}{n-2}z)^2+\frac{z^n}{n}\right]
e^{-i\frac{4\pi n}{n-2}t},
\ea
\eeq
no Hamiltonian structure is instead known.
Only after fixing the energy level, one obtains the Hamiltonian interpretation contained in the Remark 1.

\vskip 5pt
\noindent
{\bf Remark 3.} The idea that the integrability or the non-integrability of a dynamical system be
intimately related to the analyticity properties of its solutions in the complex time plane
goes back to Jacobi, Kowalewski, Painlev\'e and other eminent mathematicians. For
Kowalewski, Painlev\'e and his school, the request that the only movable singularities of ODEs
be poles has been a very successfull tool to isolate and classify important integrable cases
\cite{KPG}. The idea that integrability is compatible even with the presence of
movable branch point singularities, provided that they are not dense in the complex time plane,
is more recent, and one of us (PMS) heard several illuminating lectures on this topic
by Kruskal, more than 20 years ago. The works \cite{KC,Others} are some of the contributions
connected to this fascinating idea.

As we shall see in the following sections, the Newtonian systems
(\ref{ODEs12}) are just distinguished examples of ODEs whose solutions exhibit branch point singularities
which are everywhere dense in the complex time plane. Therefore the results presented in this
paper can also be viewed as a contribution along this line of thinking.

The paper is organized as follows. In \S 2 we discuss the topological properties of
the Riemann surfaces associated with the quadratures (\ref{quadrature2}),(\ref{quadrature3}),
for a generic $n\in\ZZ$. In \S 3 we concentrate on the physically relevant subcase in which the degree
$n$ of the monomial
is even, restricting further our analysis to a suitably factorized Riemann surface. In \S 4 we
discuss the main properties of the rectilinear and cyclic motions on such
factorized surfaces, as the degree $n$ of the monomial varies.
In \S 5 we associate with the cyclic motion the 1-parameter family of
2-dimensional mappings describing the dynamics of the center of the circle; such mappings show
typical multi-fractal behavior with periodicity islands. In \S 6 we briefly comment on how the
complexification procedures discussed in this paper can be a distinguished way of generating
dynamics exhibiting a high degree of complexity and, at the same time, amenable in
principle to analytical treatments.

\section {Riemann surfaces associated with the Newtonian dynamics: generic case}

In this section we introduce a description of the Riemann surfaces
associated with the quadratures (\ref{quadrature2})-(\ref{quadrature3}). This description is different
from that used in \cite{FG}.

\vskip 5pt
\noindent
$\mathbf {n>0}$. In the case $n>0$ we use the following representation for the surface $\Gamma$. Due
to the rotation symmetry of $\Gamma$, it is convenient to draw the cuts in the $w$-plane
as rays connecting the branch points (the $n$ roots of unity) to infinity.
Then we have 2 sheets and the map (\ref{quadrature2}) maps each of them to
a regular $n$-gone (see Fig 1):
\begin{center}
\mbox{ \epsfxsize=10cm \epsffile{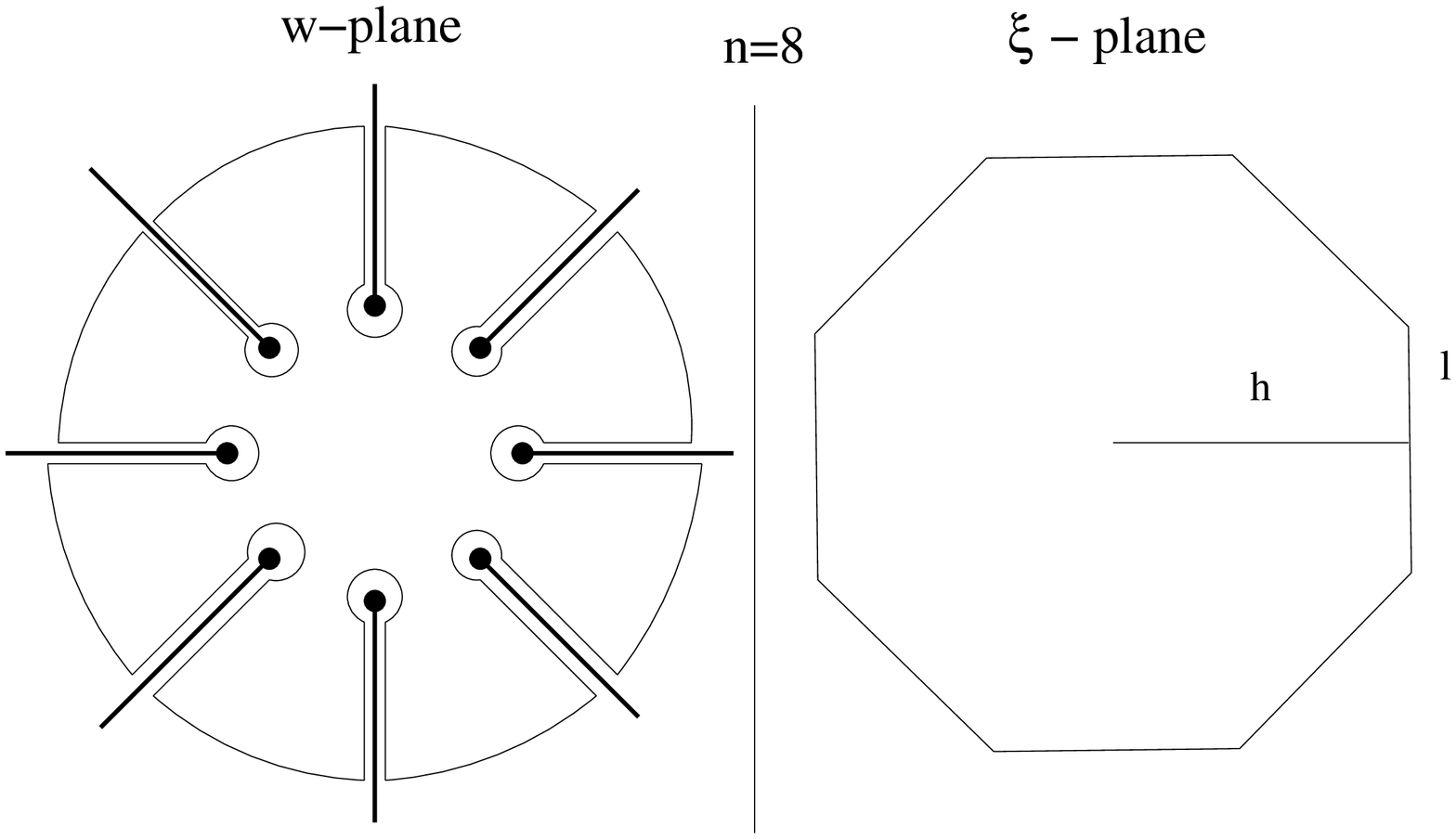}}

Fig 1.
\end{center}

Therefore the whole Riemannn surface is represented by the union of 2 regular $n$-gones, and each side
of the first $n$-gone is glued to the opposite side of the second one. Equivalently,
we can attach one side of the first $n$-gone to the corresponding side of the second
one. As the result of the above operation, we obtain a $(2n-2)$-gone, and
$\Gamma$ is obtained by gluing the opposite sides (see Fig 2).

\begin{center}
\mbox{ \epsfxsize=10cm \epsffile{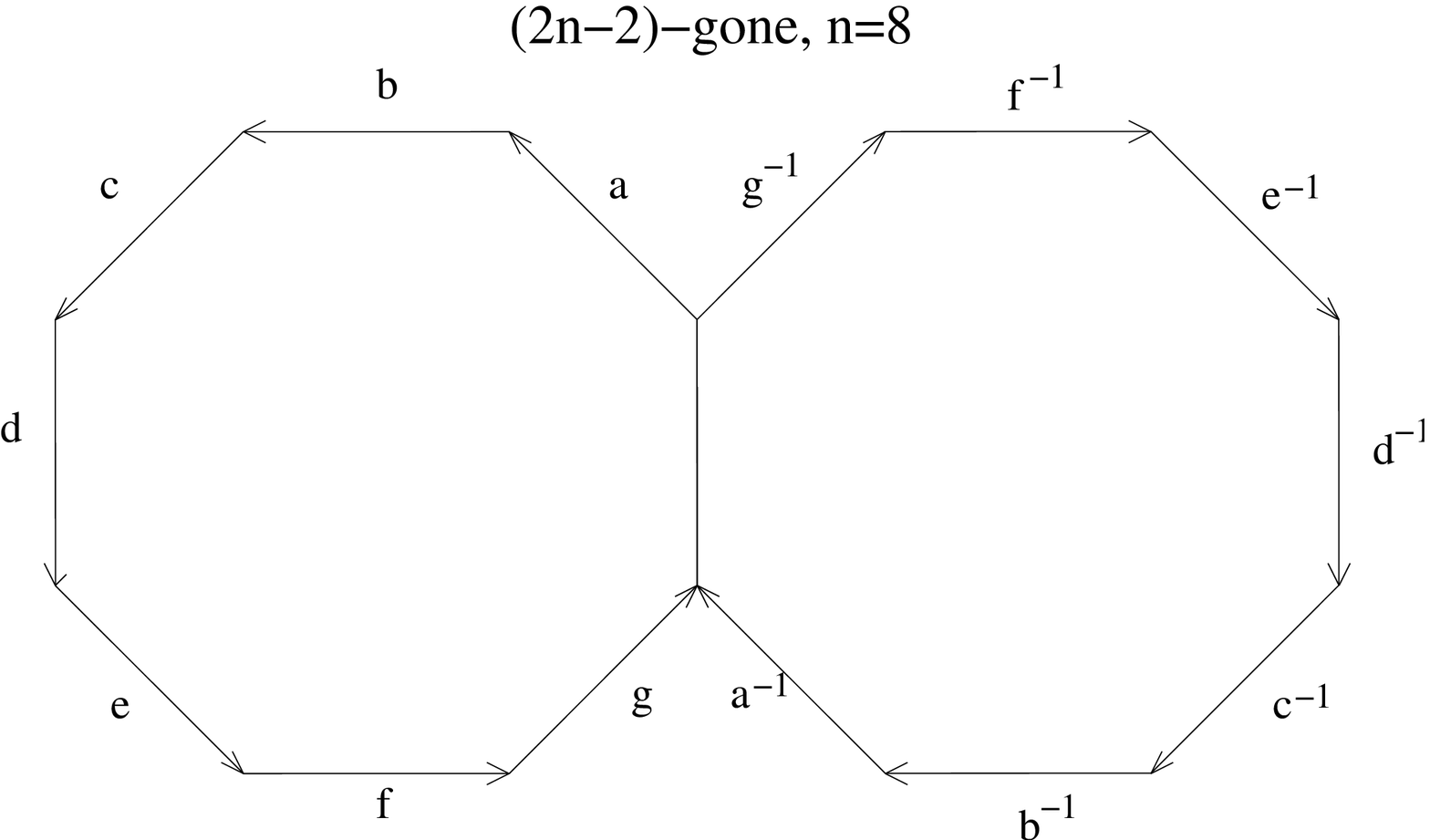}}

Fig 2.
\end{center}

Let $g(\Gamma)$ be
the genus of $\Gamma$; then, for both $n=2k$ and $n=2k-1$, $g(\Gamma)=k-1$.
The length of each side is equal to
\beq
\label{length1}
l=2\int\limits_1^{\infty}\frac{dw}{\sqrt{{w}^n-1}}
\eeq
and the distance $h$ from the center to the sides satisfies the relation $l=2h\tan(\pi/n)$.

If $n$ is even, it is also convenient to define the factorized Riemann surface
$\tilde\Gamma$, obtained by the factorization
\beq
w\rightarrow -w, \ \ \mu\rightarrow -\mu.
\eeq
$\tilde\Gamma$ can be naturally treated as the result of gluing the opposite sides
of a single regular $n$-gone \cite{DFN}. If $n=4k$ or $n=4k+2$, then
$g(\tilde\Gamma)=k$.

\vskip 12pt
\noindent$\mathbf{n<0}$. In the case $n<0$ it is convenient to draw the cuts as
rays connecting the roots of unity
to 0. If $n$ is even: $n=2k$, we have 2 sheets and the map (\ref{quadrature3}) maps each
of them to the exterior (complement) of the regular $|n|$-gone. Therefore the whole
Riemannn surface  $\Gamma$ is represented by the union of the complements to 2 regular $|n|$-gones,
and each side of the first $|n|$-gone is glued to the opposite side of the second one.
Topologically, the exterior of an $n$-gone with the infinite point added is an $n$-gone;
therefore $\Gamma$ can be represented again as the result of gluing the opposite sides of
a $(2n-2)$-gone.

If $n$ is odd: $n=2k+1$, we have one sheet and the map (\ref{quadrature3}) maps it to the
exterior of a regular $(2|n|)$-gone, which is drawn on the Riemann surface of the function
$\sqrt{w}$. The points of a side in one sheet are glued to the corresponding points of the
side in the second sheet.

\vskip 12pt
\noindent{\bf Cycles.}
In our situation it is more convenient to use, instead of the canonical basis of cycles, a more
symmetric one. Since $\Gamma$ is obtained by gluing the opposite sides of a $(2n-2)$-gone,
let us choose the center of this $(2n-2)$-gone as the starting point and denote by $C_j$ the
cycles connecting the sides $s_j$ and $s_{j+n-1}$, $j=1,\ldots,n-1$ (see Fig 3).
\begin{center}
\mbox{ \epsfxsize=10cm \epsffile{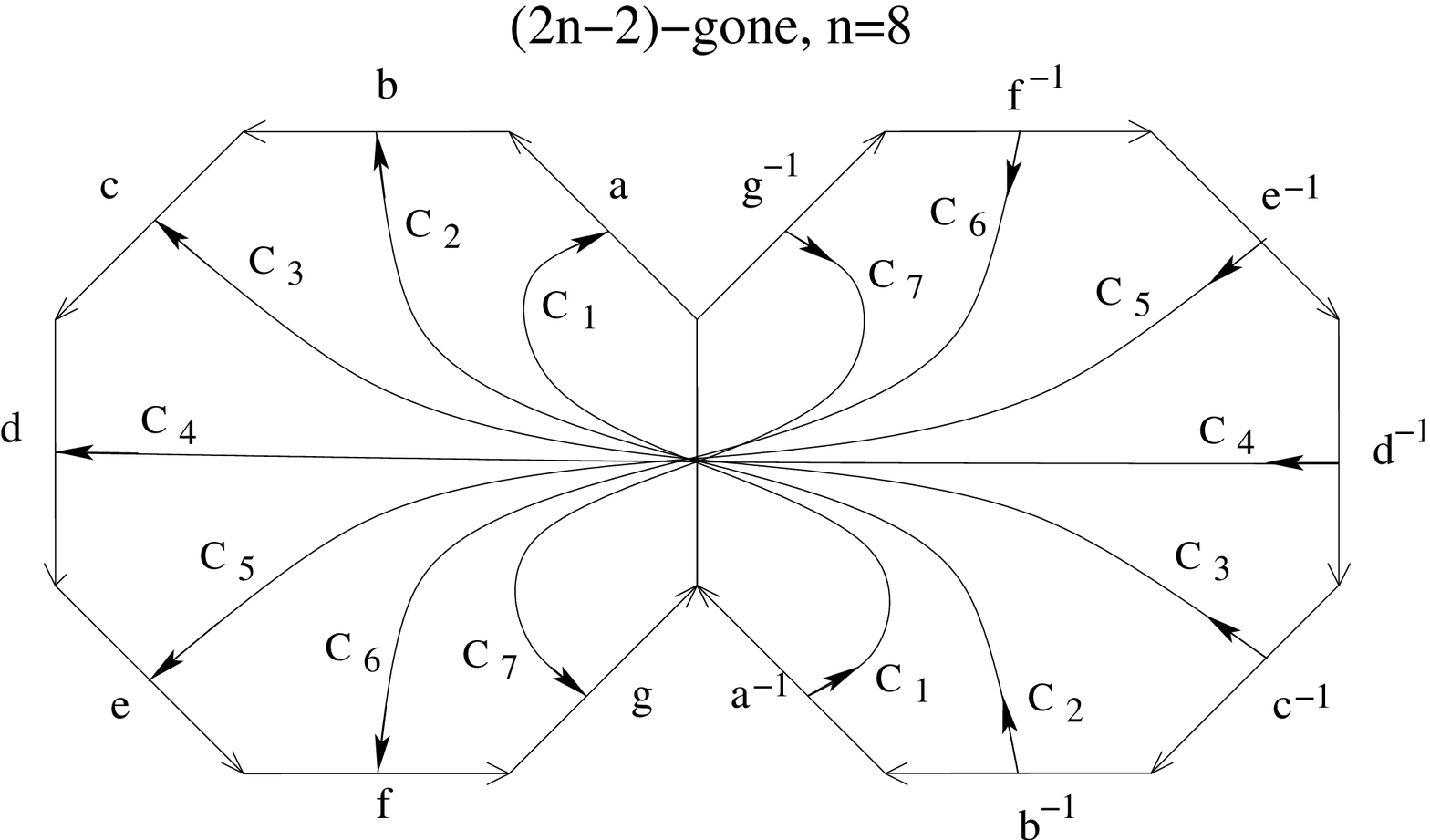}}

Fig 3.
\end{center}

If $(n-1)$ is even, these cycles
are homologically independent, $C_i\cdot C_j=1$ for $i<j$. The fundamental group of $\Gamma$
is generated by the $C_j$ with the relation $C_1C_2^{-1}C_3C_4^{-1}\ldots
C_{n-1}^{-1} C_1^{-1}C_2C_3^{-1}C_4\ldots C_{n-1}=1$.

If $(n-1)$ is odd, these cycles are dependent. It is convenient to write the fundamental group
of $\Gamma$ as the group generated by  $C_j$, $j=1,\ldots,k$ with two relations:
$C_1C_2^{-1}C_3C_4^{-1}\ldots C_{n-1}=1$, $C_1^{-1}C_2C_3^{-1}C_4\ldots C_{n-1}^{-1}=1$. In the
homological group these two relations are equivalent and they imply that
\beq
\label{cycles_relation}
C_1-C_2+C_3-C_4\ldots +C_{n-1}=0.
\eeq

As usual, the multivaluedness of $\xi(w)$, associated with the above cycles, implies that
the inverse function $w(\xi)$ be multiperiodic. Moreover, the vertices of the $n$-gone are
the only singularities of  $w(\xi)$. If $n=3,4$, this singularities are poles and
$w(\xi)$ is single-valued.
Otherwise, they are branch points $\xi_b$, and $w(\xi)$ is unbounded near them. For $n>0$
\beq
\label{BPs}
w\sim c(\xi-\xi_b)^{-\frac{2}{n-2}}, \ \ \xi\sim\xi_b.
\eeq

Due to the above multiperiodicity, these singularities are everywhere dense in the complex plane
for $n\ne 3,4,6$. These exceptional cases correspond to the equilateral triangle, the square and
the regular hexagon, which are well-known to cover exactly the plane through the
Schwarz reflection principle.

\vskip 12pt
\noindent{\bf Coverings.}
The inversion of the hyperelliptic integrals leads us to the study of the motion not only
on the surface $\Gamma$, but also on a proper covering space $\hat\Gamma$. It is well-known
that there is a one-to-one correspondence between the subgoups $\tilde\pi$ of the fundamental group
$\pi(\Gamma,x_0)$ and the non-ramified coverings of $\Gamma$.
Namely, the paths $\gamma\in\tilde\pi$ are exactly the loops which remain closed after being
lifted to covering space. If $\tilde\pi$ is the unit subgroup, we obtain the universal
covering, which for $g\ge2$ is isometric to the Lobachevsky plane. But, for the problem
of inverting the hyperelliptic integrals, it is much more natural to use ``smaller'' coverings:
\begin{enumerate}
\item The maximal Abelian covering (see \cite{Nov2}): $\tilde\pi$ is the commutant of $\pi$,
i.e. the subgroup generated by all elements of the type $g_1g_2g_1^{-1}g_2^{-1}$,
$g_1,g_2\in\pi(\Gamma,x_0)$. A closed path in
$\Gamma$ remains closed in the covering space iff the integrals of all holomorphic forms
(as well as the integrals of all meromorphic forms with only zero residues) along this
path are equal to 0.
\item The covering $\hat\Gamma$ associated with a given holomorphic form $\omega$:
the subgroup $\tilde\pi$ is formed by all closed paths $\gamma$ such that
$$
\oint\limits_{\gamma} \omega =0.
$$
\end{enumerate}
If all periods of $\omega$ are linearly independent over the rational numbers,
the covering defined
by $\omega$ coincides with the maximal Abelian one. But, if there is some rational dependence between the
periods, it is smaller.
\section {Riemann surfaces associated with the Newtonian dynamics: $n$ even}
For simplicity, in the rest of the paper we concentrate only on the case of even $n$: $n=2k$.
We shall also impose the restriction $n>0$. The reason is that for $n<0$ the rectilinear
motion is trivial (see below), while the cyclic motion can be obtained from the case
$n>0$ by time-reversal.

It is convenient to focus our attention on the motion on the factorized Riemann
surface  $\tilde\Gamma$. If the motion on $\tilde\Gamma$ is periodic, then the motion on $\Gamma$
is also periodic, and the two periods either coincide or the period on $\Gamma$ is twice bigger,
depending on whether the number of intersections with the sides is even or odd.
Therefore the motion on $\tilde\Gamma$ contains all the essential informations.
The corresponding cycles are drawn in Fig 4.
\begin{center}
\mbox{ \epsfxsize=6cm \epsffile{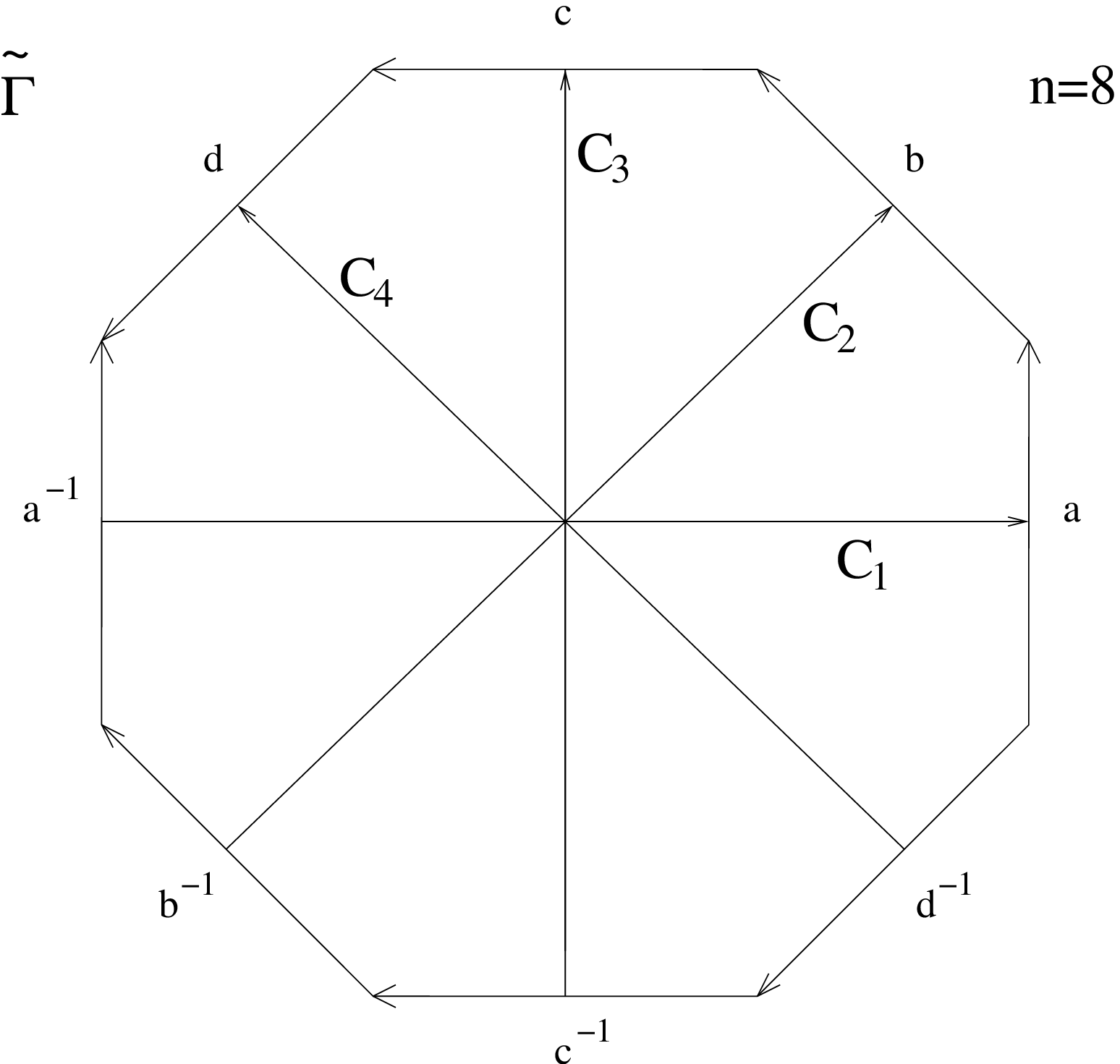}}

Fig 4.
\end{center}

Since our experiments cover the polygons with even $n$, from $n=8$ to $n=16$, let us
discuss the main
properties of the corresponding Riemann surfaces $\Gamma$, $\tilde\Gamma$.

If $n=4$, $\Gamma$ and $\tilde\Gamma$ are tori: $g(\Gamma)=g(\tilde\Gamma)=1$,
and the universal covering
coincides with the maximal Abelian covering and is exactly the complex plane. The basic
periods are $2h$ and $2ih$, and they are independent.

If $n=6$, $\tilde\Gamma$ is again a torus, $g(\tilde\Gamma)=1$,  $g(\Gamma)=2$, and $\Gamma$ is
a ramified covering of $\tilde\Gamma$ with 2 branch points. The periods along the cycles
$C_1$, $C_2$, $C_3$ are $2h$, $2e^{i\pi/3}h$, $2e^{2i\pi/3}h$ respectively, and their alternated sum is
equal to 0. Let us recall that, homologically, $C_1-C_2+C_3=0$. The covering $\hat\Gamma$ corresponding
to our integral coincides with the maximal Abelian covering.

If $n=8$, $g(\tilde\Gamma)=2$, $g(\Gamma)=3$ and $\Gamma$ is a non-ramified covering of $\tilde\Gamma$.
The basic periods of $\tilde\Gamma$ are $2h$, $2e^{i\pi/4}h$, $2e^{i\pi/2}h$, $2e^{3i\pi/4}h$, and they are
independent over the rational numbers. Therefore the covering $\hat\Gamma$ associated with our integral
again coincides  with the maximal Abelian covering.

If $n=10$, $g(\tilde\Gamma)=2$, $g(\Gamma)=4$ and $\Gamma$ is a ramified covering of $\tilde\Gamma$ with 2
branch points. The basic periods of $\tilde\Gamma$ are $2h$, $2e^{i\pi/5}h$, $2e^{2i\pi/5}h$, $2e^{3\pi/5}h$,
$2e^{4i\pi/5}h$, and the basic relation is $C_1-C_2+C_3-C_5+C_5=0$. Again the covering $\hat\Gamma$
associated with our integral coincides with the maximal Abelian covering.

If  $n=12$, $g(\tilde\Gamma)=3$, $g(\Gamma)=5$ and $\Gamma$ is a non-ramified covering of $\tilde\Gamma$.
The basic periods of $\tilde\Gamma$ are $2h$, $2e^{i\pi/6}h$, $2e^{i\pi/3}h$, $2e^{i\pi/2}h$, $2e^{2i\pi/3}h$,
$2e^{5i\pi/6}h$; therefore
\beq
\label{extra12}
\oint\limits_{C_1-C_3+C_5} \!\!\!\!\!\!\!\omega = \oint\limits_{C_2-C_4+C_6} \!\!\!\!\!\!\!\omega = 0.
\eeq
It is the first case in which we have nontrivial integer relations between the periods:
we have 6 independent directions in the universal covering, but only 4 of them remain
independent on the covering $\hat\Gamma$ associated with our hyperelliptic integral.

If $n=14$,  $g(\tilde\Gamma)=3$, $g(\Gamma)=6$ and $\Gamma$ is a ramified covering of $\tilde\Gamma$ with 2
branch points. The only relation between the periods is the one prescribed by the homological group:
$C_1-C_2+C_3-C_4+C_5=0$. The covering $\hat\Gamma$ associated  with our integral coincides with
the maximal Abelian covering.

The last case studied in our numerical experiments is $n=16$. In this case $g(\tilde\Gamma)=4$,
$g(\Gamma)=7$ and $\Gamma$ is a non-ramified covering of $\tilde\Gamma$. The basic periods
of $\tilde\Gamma$ are independent over the rational numbers, and we have the full Abel covering.

The points of the maximal Abelian covering are represented by the points of the basic polygon plus
the total shift
\beq
\label{total_shift}
{\cal T}=\sum\limits_{j=1}^{k} m_j C_j,
\eeq
where $m_j$ is the number of crossings of the trajectory with the side $s_j$ minus the number of crossings
with the opposite side $s_{j+k}$. For odd $k$, the relation (\ref{cycles_relation}) should be taken into
account. The simplest way is to substitute
$C_k = - C_1 + C_2 + \ldots +C_{k-1}$ into (\ref{total_shift}), obtaining
\beq
\label{total_shift2}
{\cal T}=\sum\limits_{j=1}^{k-1} (m_j+(-1)^j m_k) C_j.
\eeq

Due to the rotation symmetry of $\tilde\Gamma$, we can have extra integer relations among the
periods of the form. For instance, if $n=4lk'$, where $l$ is odd, the cycles $C_m$, $C_{m+2k'}$,
$C_{m+4k'}$, \ldots $C_{m+2k'(l-1)}$ are independent, but the corresponding periods satisfy the
relations:
\beq
\label{extra_gen}
\oint\limits_{C_m-C_{m+2k'}+C_{m+4k'}+\ldots+  C_{m+2k'(l-1)}} \!\!\!\!\!\!\! \!\!\!\!\!\!\! \!\!\!\!\!\!\!
\!\!\!\!\!\!\!
\omega = 0, \ \ m=1,2\ldots,2k';
\eeq
see, for instance, (\ref{extra12}) for  $n=12$.
In these cases, instead of studying the maximal Abelian covering,
we study the smaller covering $\hat\Gamma$ associated with our hyperelliptic integral,
and we have extra relations. Therefore
\beq
\label{total_shift3}
{\cal T}=\sum\limits_{j=1}^{\tilde n} \tilde m_j C_j
\eeq
for suitable $\tilde m_j$, where $\tilde n$ is the number of independent periods.
\section{Rectilinear and cyclic motions}
{\bf Rectilinear motion.}
If $n<0$ the qualitative picture is rather simple. The integrand is a meromorphic differential,
and each trajectory goes to  $\infty$ for $t\rightarrow\pm\infty$. In terms of the vector field, it
means that we have either 2 (for even $n$) or 1 (for odd $n$) fixed points, all the trajectories reach
them in an infinite time, and generic initial conditions result in regular motions without sensitive
dependence on the initial condition.

If $n>0$ we are in the situation in which the Hamiltonian flow is defined by a holomorphic form.
The basic features of such motion for $g>1$ were intensively studied by many authors (see
the literature at the end of \S 1). It
is convenient to choose a Poincar\'e cycle transversal to the flow, and study the corresponding
Poincar\'e map. For a given trajectory, the number $N$ of intersections with the Poincar\'e cycle
is used as the discrete time.

For generic data (a generic algebraic Riemann surface and a generic holomorphic form)
the motion is aperiodic, the image on the maximal Abelian covering has a distinguished
direction $\vec v_0$ (which is Poincar\'e dual to the cohomological class of the form). In our
case
\beq
\vec v_0 =\sum\limits_{j=1}^k (v_0)_j C_j
\eeq
where
\beq
(v_0)_j = (\vec\alpha, \hat n_j), \ \ \vec\alpha=(\Re\alpha,\Im\alpha),
\hat n_j = (\cos(2\pi(j-1)/n, \sin(2\pi(j-1)/n)),
\eeq
$\hat n_j$ denotes the normal vector to the side $s_j$. Let us recall that, if $n=2k$ and $k$
is odd, the vectors $C_j$ are dependent, and  for the maximal Abelian covering we can use the basis
$C_1$, \ldots, $C_{k-1}$. Then
\beq
\vec v_0 =\sum\limits_{j=1}^{k-1} (\tilde v_0)_j C_j, \ \ (\tilde v_0)_j =(v_0)_j+ (-1)^j (v_0)_k.
\eeq
If the number $N$ of iterations is sufficiently large, the total shift from the starting point is described
by the formula
\beq
\label{lyapunov1}
{\cal T}=Nc\vec v_0 + o(N),
\eeq
where $c$ is an order 1 constant.
The Poincar\'e map is ergodic, and this implies that, for any pair of arbitrary close initial points,
the stripe between their trajectories sooner or later meets a branch point, located at the vertex
of the polygon. After this event the trajectories separate to a distance of order 1. The difference
between these trajectories is encoded in the $o(N)$ term of the asymptotics. The corrections to
the leading asymptotic term were studied in \cite{Z2}, \cite{Z3} (see also the review \cite{Z1}),
 where the following statements were proved:

\begin{enumerate}
\item Let us consider the projection $P_{\vec v_0}$ along the vector $\vec v_0$ to the
complementary space. Then there exists a vector $\vec v_1$ in this complementary space
such that the projection has the form:
\beq
\label{lyapunov2}
P_{\vec v_0} {\cal T} = N^{\lambda_1} c_1(N) \vec v_1 +  o(N^{\lambda_1}), \ \ 0<\lambda_1<1.
\eeq
where $c_1(N)$ is a random function of $N$ of order 1
\item  For any given starting point, it is possible to define a special subsequence $\{N_l\}$
such that, going from $N_l$ to $N_{l+1}$, the trajectory returns much closer to the starting point
(for detailes see \cite{Z1}). Using this subsequence, it is possible to continue the above
procedure, defining the vectors $\vec v_2$, \ldots, $\vec v_{g-1}$ and the sequence
$\lambda_2$, \ldots,  $\lambda_{g-1}$ such that $1>\lambda_1>\lambda_2\ldots\lambda_{g-1}>0$.
\item If $\circ$ denotes the intersection form on cycles, it generates a symplectic scalar
product on vectors. In this symplectic product $\vec v_i\circ\vec v_j=0$ for all $i,j$.
\end{enumerate}

If we consider a pair of close trajectories, after a sufficiently large number of iterations
they deviate on the maximal Abelian covering, and, according to the formula (\ref{lyapunov2}), the distance
on the maximal Abelian covering increases as $N^{\lambda_1}$. Therefore $\lambda_1$
can be interpreted as the main
Lyapunov exponent, and the remaining $\lambda_j$'s can be interpreted as the other Lyapunov
exponents (see \cite{Z1}). Additional information about the numerical values of these exponents
for small $g$ as well as some analytic results can be found in the review \cite{Z1}.

Formally, the results of the papers \cite{Z1}-\cite{Z3} are valid for generic Riemann surfaces,
while our polygon is highly symmetric. But numerical experiments show good agreement with this theory.

\vskip 12pt
\noindent{\bf Cyclic motion.} As far as we know, the study of cyclic motions on Riemann surfaces
began only recently (see the literature at the end of \S 1) and the picture is much less clear. Here
we present the results of numerical experiments proceeded by the
authors, and we give some preliminary qualitative explanation of the observed phenomena, based
on the topological structure of the surface.

We always assume that our trajectories are generic in the following sense: they never hit the
vertices of the polygon, otherwise the motion becomes undefined.

For the numerical simulations we used the following technique. As it was mentioned above, we make
our calculations on $\tilde\Gamma$, which is represented by gluing the sides of the regular $n$-gone,
$n=2k$. In fact, to recover the dynamics on $\Gamma$ for even $n$,
it is sufficient to study the dynamics on $\tilde\Gamma$ and check at each step if the number
of intersections of our trajectory with the sides of the polygon is even or odd. If it is even,
the trajectory is on the same sheet, and if it is odd, the trajectory is on the opposite sheet.

Any time our trajectory crosses the side of the polygon, it is shifted of the corresponding
period.

In the cyclic case the trajectory is periodic iff, at some iteration, the total shift on
the covering associated with the quadrature becomes zero.

We first discuss the exceptional cases. If $n=4$, $\Gamma$ and $\tilde\Gamma$ are tori,
$w(\xi)$ is single-valued and the motion is always periodic with period 1; i.e. it is isochronous.
If $n=6$, $\Gamma$ is a two-sheeted ramified covering of the torus $\tilde\Gamma$ with two branch
points. Therefore the motion is always periodic, with period 1 or 2, depending on  whether the number
of square root type branch points inside the trajectory is even or odd. These two cases have been
already considered in \cite{FG}.

We now discuss the non-exceptional cases.

For given $n=2k$, if the diameter $D$ of the circle is smaller than $l$, where $l$ is the side length,
the motion is periodic, and there is at most 1 branch point inside the circle. If there is
no branch point inside, the period is 1 and the motion is isochronous; otherwise it is equal
to $(n-2)/2$ for even $k$ and to $(n-2)/4$ for odd $k$.

For bigger cycles the
situation becomes much more complicated, and we studied it using a series of computer experiments,
which have not exhausted the whole phase space. These experiments show the following.

\begin{enumerate}
\item The period is very sensitive to the radius $R$ and to the position $\xi_c$ of the center
of the circle.

\item Our observation shows an essential difference between the cases $n=8,10,12$ and $n=14,16$.

Let us discuss the first group $n=8,10,12$. From the numerical experiments it seems
that, for data of full measure, the motion is periodic, and the typical periods grow
exponentially with the ratio $D/l$. We have observed periods of order $10^8$
for $D/l\sim 20$. For bigger $D$, we have observed starting points for which the period was not
found after $2\cdot 10^9$ iterations.

For $n=14$ and $n=16$ we have found that, for many initial configurations starting from
$D/l\sim 2$, the period was not found after $2\cdot 10^9$ iterations. Our conjecture is that,
in this cases, the aperiodic trajectories are non-exceptional.

\item If we consider the trajectories for which the period was either not found or was sufficiently
large, the total shift on the covering space seems to exhibit the following behaviour:
\beq
\label{root_law}
{\cal T} \sim N^{\lambda} \vec v(N), \ \ \lambda\sim \frac12,
\eeq
where $N$ is the large number of iterations and $\vec v(N)$ is a random vector of order 1.
Therefore, the distance on $\hat\Gamma$ of two close trajectories increases approximately as
$N^{\lambda}$, and $\lambda\sim\frac12$ can be interpeted as the principal Lyapunov exponent.
\end{enumerate}

Formula (\ref{root_law}) suggests that, for sufficiently large $N$, the shift ${\cal T}$ can be
approximated by the random motion on a lattice. Now we show that, accepting this assumption,
one can explain the difference between the above two classes.

Let us stress that, for cyclic motions, the total shift in the covering space can be arbitrarily large
but, after being projected to the time-plane, the total shift is bounded by the condition:
\beq
\label{principle_inequality}
|\sum\limits_{j=1}^{k} m_j e^{(j-1)\pi/k}| < \frac{D}{2h} + 1/\cos(\pi/n).
\eeq

This inequality imposes a severe constraint on the vector $(m_1,\ldots,m_k)$. Let us denote by
``effective dimension'' the maximal number of $m_j$'s which can be chosen arbitrarily.
If these components are fixed, there is only a finite number of options for the other components.

Let us point out that the effective dimension is equal to the number of independent shifts in the
covering $\hat\Gamma$ associated with the quadrature minus 2. Indeed, if we fix any pair of basic shifts,
say $e^{(j-1)\pi/k}$, $j=1,2$,  which form a basis in the complex plane over $\mathbb R$, and we assign
arbitrary integer values to all the other $m_j$'s ($j\ne 1,2$), then the two coefficients $m_1$, $m_2$
are constrained by the inequality (\ref{principle_inequality}) to take only a finite number of values.

From the considerations made in Setion 3, it follows that, for $n=8,10,12$ the effective dimension
is 2, while for $n=14$ it is equal to 4  and  for $n=16$ it is equal to 6.

On the basis of the numerical experiments we assume that, for sufficiently large $N$,
our dynamics be approximated  by a random motion in the ``lattice of effective dimension''.
It is well-known that, for the 2-d symmetric random motion, the probability to return to the starting
point is equal to 1; therefore we expect that, with probability 1, this motion be periodic. If the
dimension is higher than 2, with non-zero probability we do not return to the starting point,
and there are aperiodic trajectories. Therefore the random hypothesis implies that,
for  $n=8,10,12$ (cases in which the effective dimension is $2$) the typical motion is periodic,
while  for  $n=14$ and $n=16$ (cases in which the effective dimension is greater than $2$) there are
aperiodic domains of positive measure, in agreement with our observations.

\section{The center map as a fractal}
In the rectilinear motion it is convenient to represent the
dynamics on the surface in terms of the Poincar\'e map on a
transversal cycle. It is the well-known ``interval exchange map''.
In the cyclic motion it is convenient to study the dynamics of the
center $\xi_c$ of the cycle, for any fixed radius $R$.

The dynamical rule is the following: any time the circle crosses a side $s_j$ of the $n$-gone,
the center is shifted in the following way:
\beq
\xi_c\rightarrow \xi_c - 2h e^{2\pi(j-1)/n}.
\eeq
It is a 1-parameter family of deterministic 2-dimensional mappings (the arbitrary parameter being
the radius of the circle). We call this mapping the ``center map''.

It is interesting to study the center map for aperiodic trajectories (of course, up to the
computer experiment limits). Considering as basic illustrative example
the polygon $n=16$, $h=1$ centered at the origin, we have made a series of
experiments showing interesting generic multi-fractal behaviour with periodicity islands.

In the first set of experiments we fixed $R=0.358$ and we varied the initial position of the
center. In Fig 5 we choose $\xi_c= (1.1610,-0.4037)$, obtaining the following fractal:

\begin{center}
\mbox{ \epsfxsize=13.5cm \epsffile{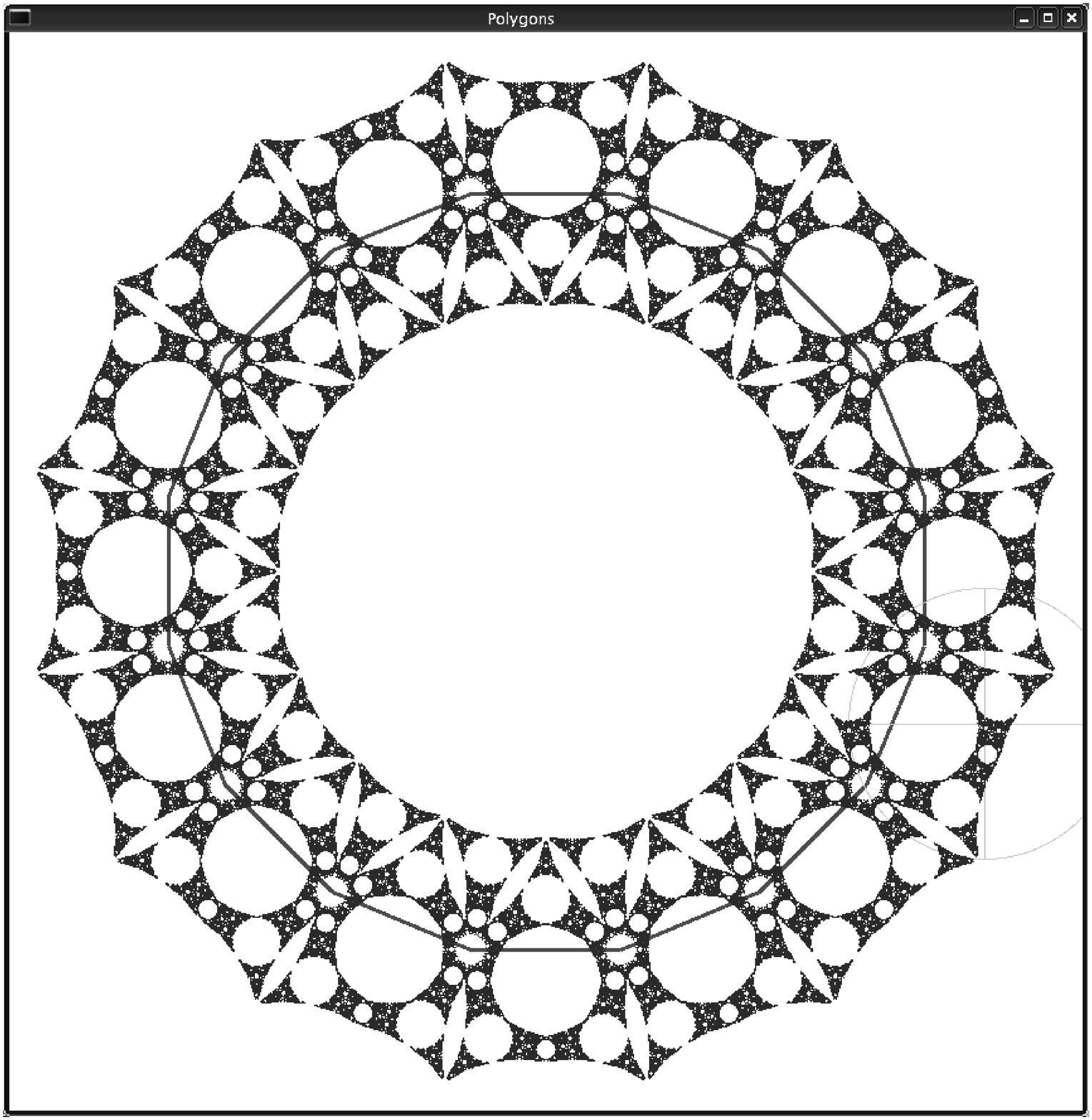} }
Fig 5.
\end{center}

In this figure, together with the fractal, we see also the 16-gone and the initial circle,
located in the south-east
position (it is drawn by a very thin line). The initial center $\xi_c$ is marked by the intersection
of the inner lines. We remark that this fractal shows the rotation symmetry of the polygone.

In Fig 6 we see the same fractal, but magnified 33 times. The initial center is marked again
by the intersections of the inner lines.

\begin{center}
\mbox{ \epsfxsize=14cm \epsffile{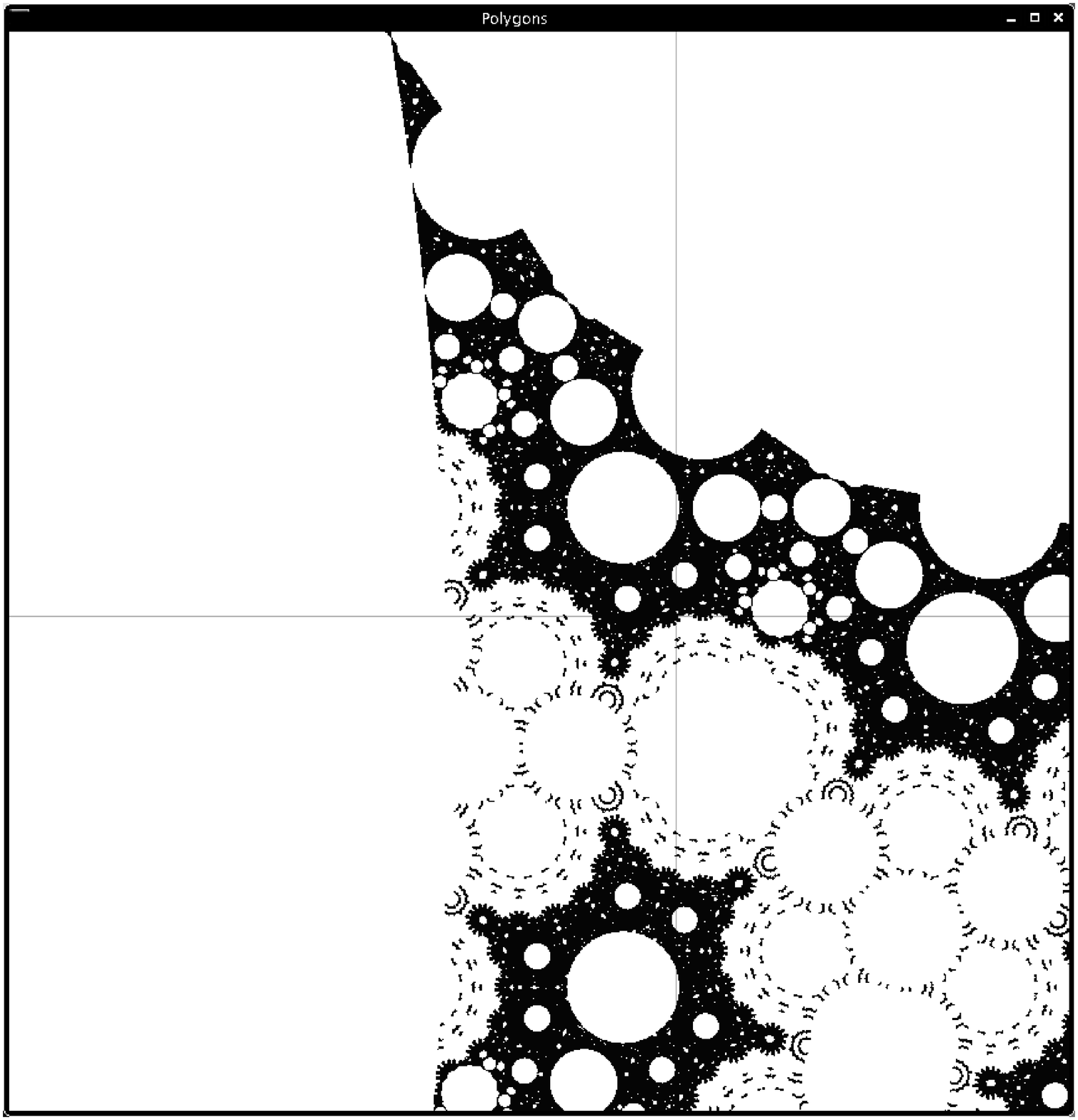} }

Fig 6.
\end{center}

This figure makes clear the fractal structure. It turns out that this fractal is the boundary
of some islands inside which the center map is periodic with fixed periods.
Typically the period increases as the size of the island decreases.

Choosing the starting point $\xi_c$ inside the fractal region, we obtain a fractal coinciding with
this one up to the picture resolution. Choosing the starting point $\xi_c=(1.1572,-0.4151)$ ouside
the fractal, but on the boundary of the periodicity islands, we obtain a different fractal, living
in the complementary region (see Fig 7).

\begin{center}
\mbox{ \epsfxsize=13cm \epsffile{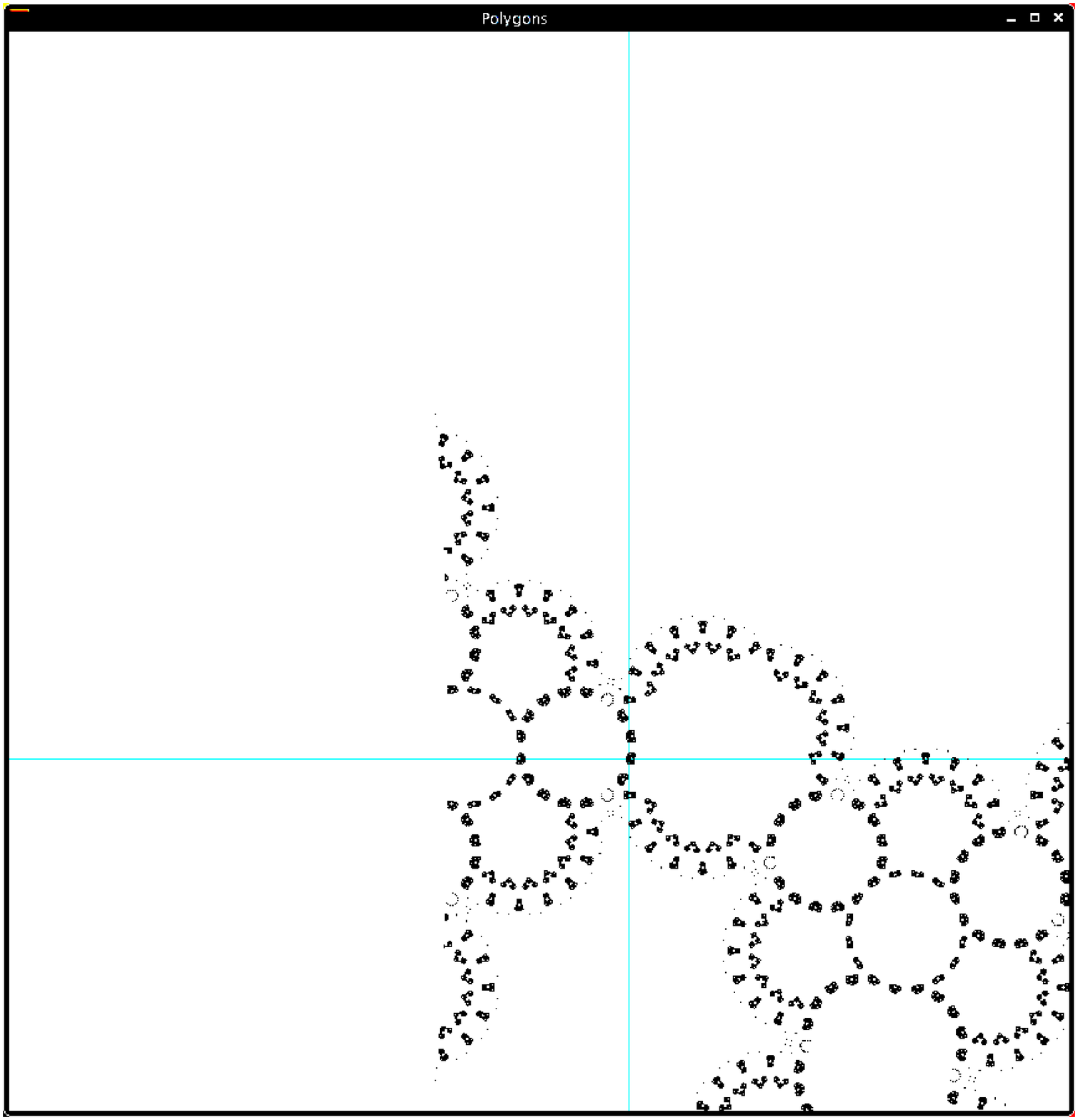} }

Fig 7.
\end{center}

In the second series of experiments we change the radius $R$.
For a sufficiently small change of radius the fractal changes, but keeping all the principal
features. Choosing, instead, a sufficiently different radius, we obtain completely
different fractal pictures. For instance, for $R=0.6$, we obtain the rich fractal
shown in Fig 8 (compare it with the fractal in Fig 5).

\begin{center}
\mbox{ \epsfxsize=14cm \epsffile{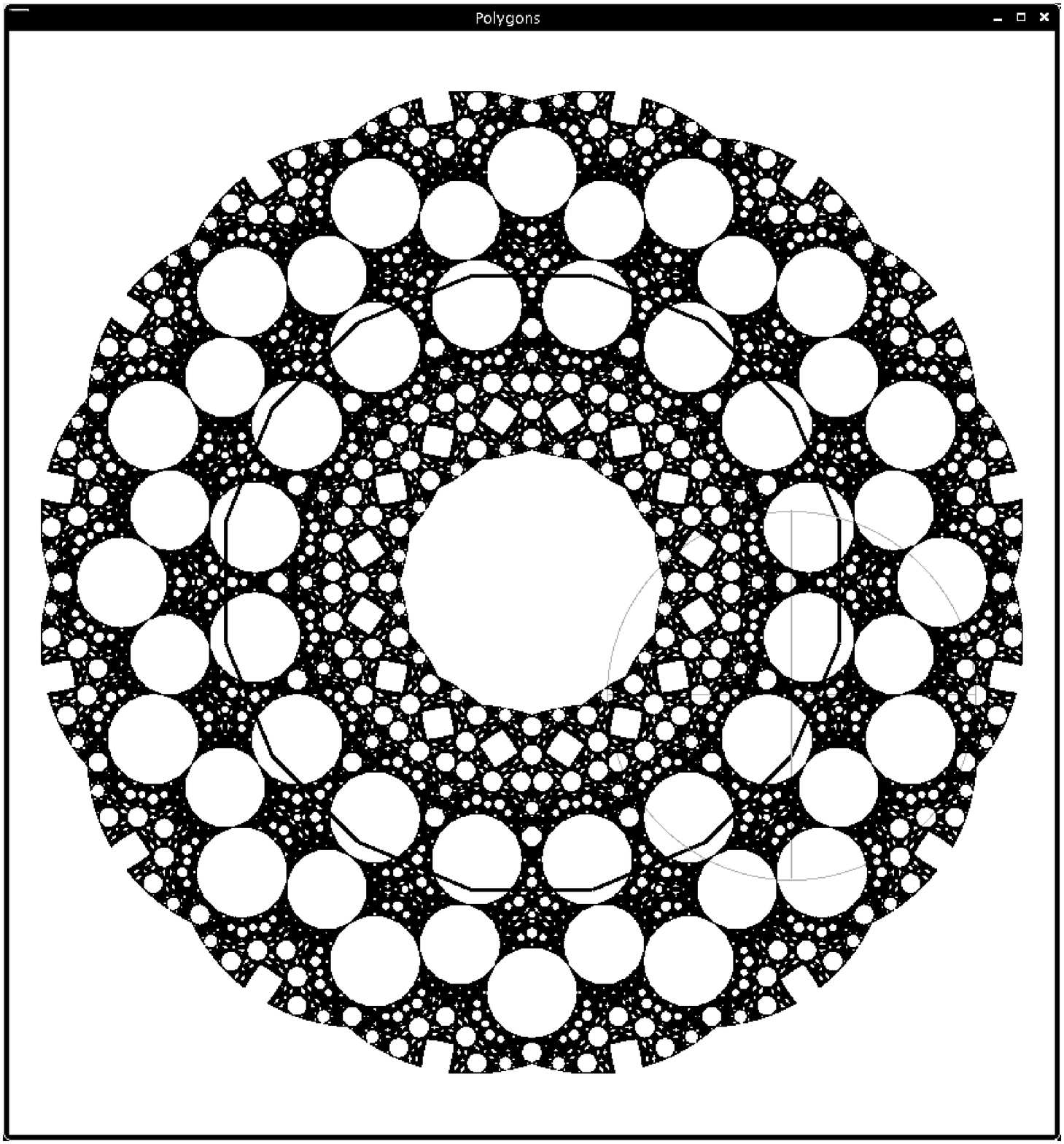} }

Fig 8.
\end{center}

\newpage

Fig 9 shows a detail of the fractal of Fig 8, magnified 10 times, containing a part of the initial
circle.

\begin{center}
\mbox{ \epsfxsize=14cm \epsffile{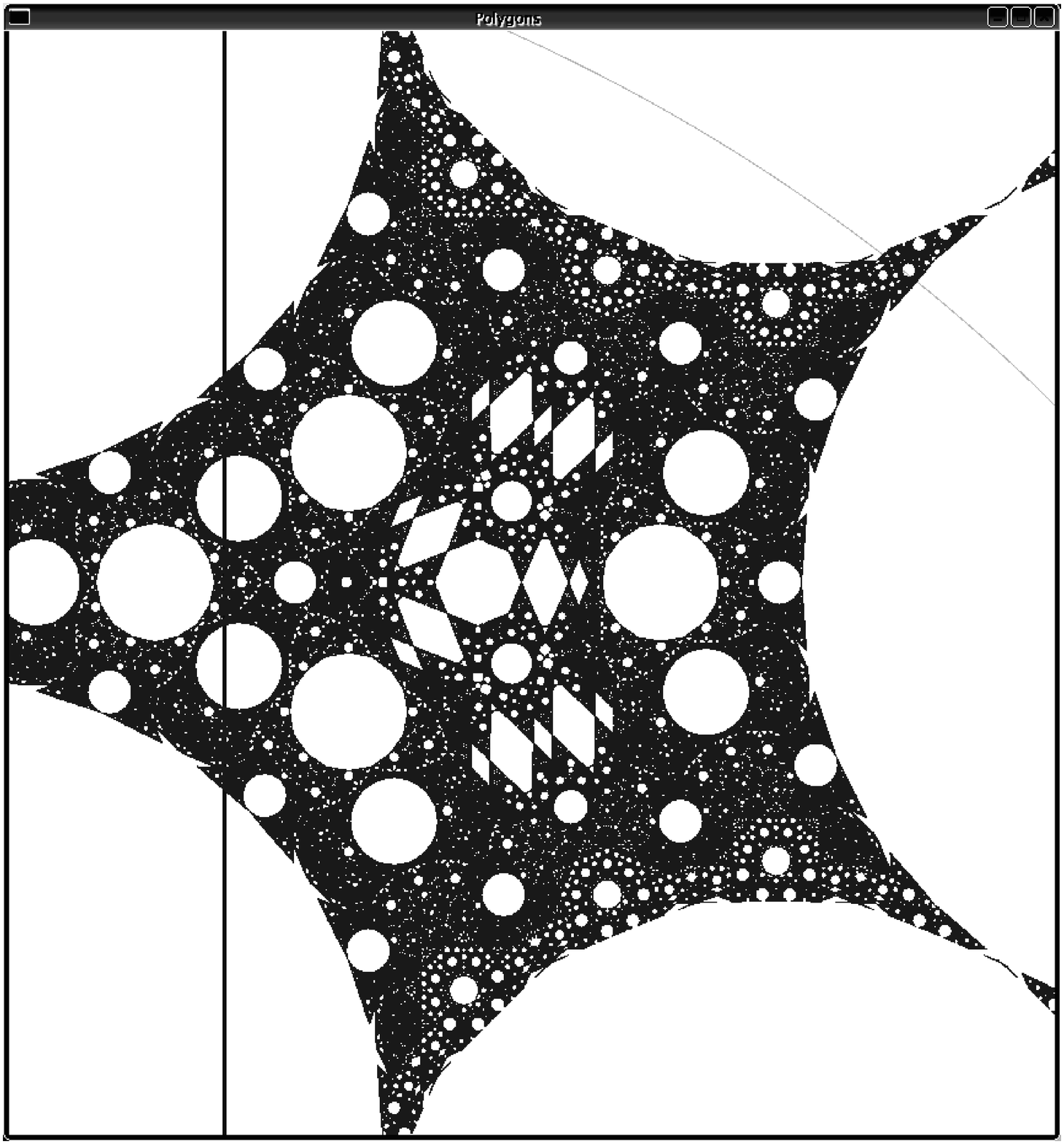} }

Fig 9.
\end{center}

\newpage

Fig 10 shows a detail of Fig 9, magnified 4 times more. These magnifications make evident
the extremely rich fractal nature of the center map.

\begin{center}
\mbox{ \epsfxsize=14cm \epsffile{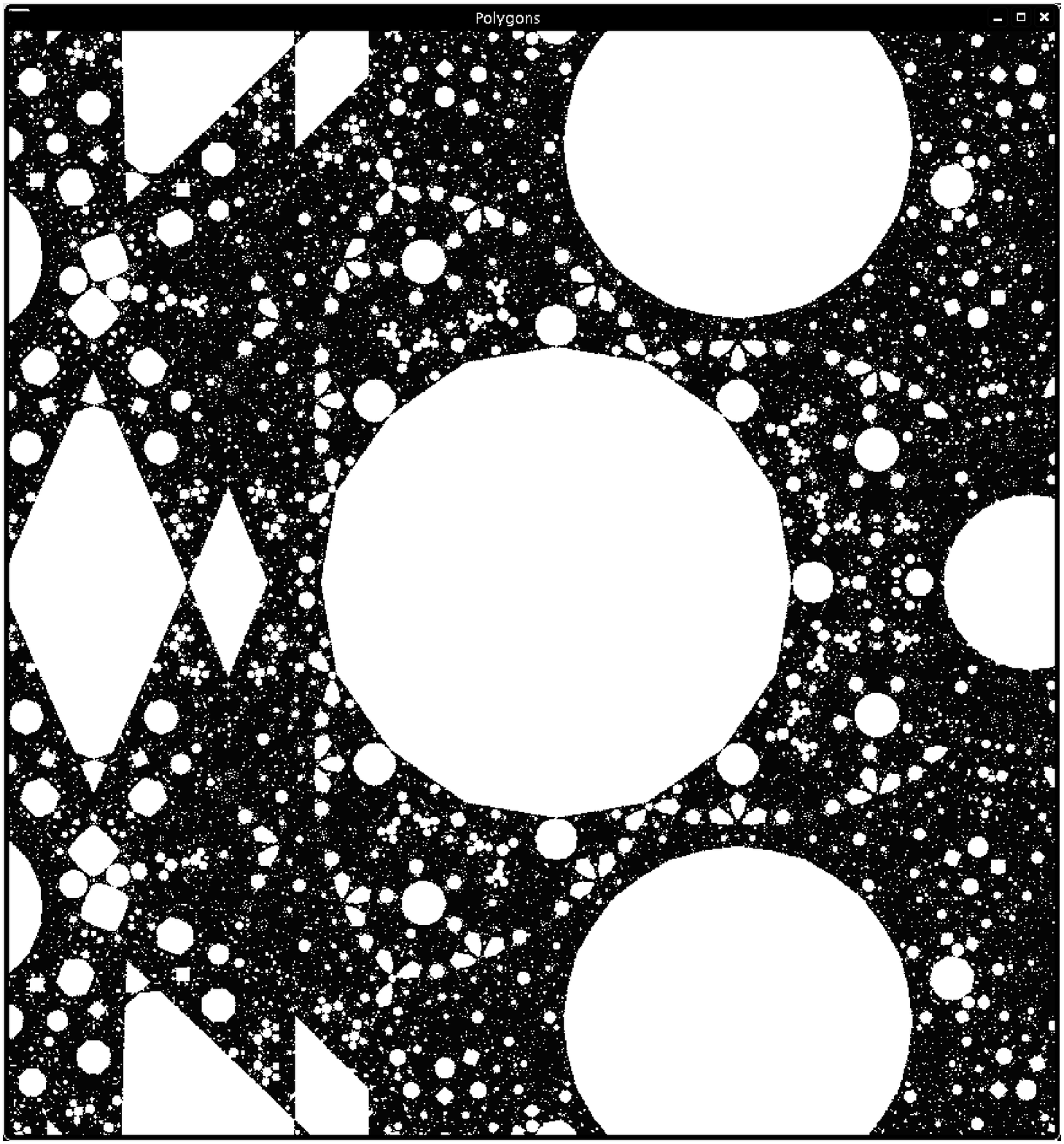} }

Fig 10.
\end{center}

\newpage

\section{Complexification and Complexity}

We end this work with a consideration. The results of this paper strongly suggest that
proper complexifications of dependent and independent variables of a given ODE are very
fruitful and rich operations, enabling one to generate complicated dynamics amenable to exact
analytical treatments. Indeed they allow to go, for instance, from the elementary one dimensional
Newtonian motion
in a monomial potential, to the complicated dynamics (\ref{ODEs12}b),
whose high degree of complexity is well illustrated by
the fractal behavior of the associated center map (see \S 5). We feel that dynamics
generated through complexification procedures could serve as prototypical examples of a new
paradigm of chaos, potentially interesting in applications.

\vskip 15pt
\noindent
{\bf Acknowledgements}.

\vskip 5pt
\noindent
This research was initiated in collaboration with F. Calogero,
who is also the initiator of the studies of dynamical systems associated with
cyclic motions on Riemann surfaces. Unfortunately other duties
prevented him from taking an active part in the development of
the results reported in this paper, and he therefore preferred not to
sign this paper, although he did partecipate in the very early stage of this
research and he constantly encouraged our investigations. It is therefore a pleasure
to us to acknowledge his important contribution, with the hope that he will join us
in the future developments of this research.

PMS acknowledges useful discussions with D. Gomez-Ullate and Y. Fedorov concerning
their results \cite{FG} on the analytical study of the dynamical systems (\ref{ODEs12}b).
These results, together with the results \cite{Calogero5,Induti} of
F. Calogero and E. Induti, have been the main motivation for the present study.
He also acknowledges the stimulating questions and remarks of the students of
the specialistic course ``Nonlinear evolutionary systems'' he gave in the University of Roma ``La Sapienza''
in the Spring 2006, in which some of the ideas and results
contained in this paper were first presented and tested.

The visit of PG to Rome was supported by the INFN grant 2006. PG was also
supported by the RFBR grant 04-01-00403a.

\end{document}